\documentclass{ws-ijmpa}
\begin{document}
\markboth{Moshe Carmeli}
{The Cosmological Constant}
\catchline{}{}{}
\title{FUNDAMENTAL APPROACH TO THE \\ COSMOLOGICAL CONSTANT ISSUE}
\author{\footnotesize MOSHE CARMELI\footnote{Email: 
carmelim@bgumail.bgu.ac.il.}}

\address{Department of Physics,  Ben Gurion University, Berr Sheva 84105,
Israel.}
\maketitle
\begin{abstract}
We use a Riemannian four-dimensional presentation of 
gravitation in which the coordinates are those of Hubble, i.e. distances and 
velocity rather than the traditional space and time. We solve the field 
equations and show
that there are three possibilities for the Universe to expand. The theory 
describes the Universe as having a three-phase evolution with a decelerating
expansion, followed by a constant and  an accelerating expansion, and it 
predicts that the Universe is now in the latter phase. It is
shown, assuming $\Omega_m=0.245$, that the time at which the Universe goes over
from a decelerating to an accelerating expansion, i.e., the constant-expansion
phase, occurs at 8.5 Gyr ago. Also, at that time the cosmic radiation 
temperature was 146K. Recent
observations of distant supernovae imply, in defiance of expectations, that 
the Universe's growth is accelerating, contrary to what has always been 
assumed, that the expansion is slowing down due to gravity. Our theory 
confirms these recent experimental results by showing that the Universe now is 
definitely in a stage of accelerating expansion. The theory predicts also that 
now there is a positive pressure, $p=0.034g/cm^2$, in the Universe. Although 
the theory has no
cosmological constant, we extract from it its equivalence and show that 
$\Lambda=1.934\times 10^{-35}s^{-2}$. This value of 
$\Lambda$ is in excellent agreement with measurements. It is 
also shown that the three-dimensional space of the Universe is Euclidean, as 
the Boomerang experiment shows. 
\end{abstract}
\section{Preliminaries}
As in classical general relativity we start our discussion in flat 
spacevelocity which
will then be generalized to curved space. 

The flat-spacevelocity cosmological metric is given by 
$$ds^2=\tau^2dv^2-\left(dx^2+dy^2+dz^2\right).\eqno(1)$$
Here $\tau$ is Hubble's time, the inverse of Hubble's constant, as given by
measurements in the limit of zero distances and thus zero gravity. As such, 
$\tau$ is a constant, in fact a universal constant (its
numerical value is given in Section 8, $\tau=12.486$Gyr). Its role 
in cosmology theory resembles that of $c$, the speed of light in vacuum, in 
ordinary special relativity. The velocity $v$ is used here in the sense of 
cosmology, as in Hubble's law, and is usually not the time-derivative of the 
distance.

The Universe expansion is obtained from the metric (1) as a null condition,
$ds=0$. Using spherical coordinates $r$, $\theta$, $\phi$ for the metric 
(1), and the fact that the Universe is spherically symmetric ($d\theta=
d\phi=0$), the null condition then yields $dr/dv=\tau$, or upon integration
and using appropriate initial conditions, gives $r=\tau v$ or $v=H_0r$, i.e.
the Hubble law in the zero-gravity limit.        

Based on the metric (1) a cosmological special relativity 
(CSR) was presented in the text \cite{1} (see Chapter 2). In this theory the receding 
velocities of galaxies and the
distances between them in the Hubble expansion are united into a 
four-dimensional pseudo-Euclidean manifold, similarly to space and time in
ordinary special relativity. The Hubble law is assumed and is written in an
invariant way that enables one to derive a four-dimensional transformation 
which is similar to the Lorentz transformation. The parameter in the new 
transformation is the ratio between the cosmic time to $\tau$ (in which the 
cosmic time is measured backward with respect to the present time). 
Accordingly, the new transformation relates physical quantities at different 
cosmic times in the limit of weak or negligible gravitation. 

The transformation between the four variables $x$, $y$, $z$, $v$ and $x'$, 
$y'$, $z'$, $v'$ (assuming $y'=y$ and $z'=z$) is given by
$$x'=\frac{x-tv}{\sqrt{1-t^2/{\tau}^2}}, \ \ \ \
v'=\frac{v-tx/{\tau}^2}{\sqrt{1-t^2/{\tau}^2}}, \ \ \ \ 
y'=y, \ z'=z.\eqno(2)$$   
Equations (2) are the {\it cosmological transformation} and very much resemble 
the
well-known Lorentz transformation. In CSR it is the relative cosmic time which
takes the role of the relative velocity in Einstein's special relativity. The
transformation (2) leaves invariant the Hubble time $\tau$, just as the
Lorentz transformation leaves invariant the speed of light in vacuum $c$. 

\section{Cosmology in Spacevelocity}
A cosmological general theory of relativity, suitable for the large-scale 
structure of the Universe, was subsequently developed \cite{2,3,4,5}. In the 
framework of cosmological general relativity (CGR) gravitation is described
by a curved four-dimensional Riemannian spacevelocity. CGR incorporates the
Hubble constant $\tau$ at the outset. The Hubble law is assumed in CGR as a
fundamental law. CGR, in essence, extends Hubble's law so as to incorporate 
gravitation in it; it is actually a {\it distribution theory} that relates distances 
and velocities between galaxies. The theory involves only measured quantities
and it takes a picture of the Universe as it is at any moment.  

The foundations of any gravitational theory are based on the principle of
equivalence and the principle of general covariance \cite{6}. These two principles lead immediately
to the realization that gravitation should be described by a four-dimensional
curved spacetime, in our theory spacevelocity, and that the field equations 
and the equations of motion should be written in a generally covariant form.
Hence these principles were adopted in CGR also. Use is made in a 
four-dimensional Riemannian manifold with a metric $g_{\mu\nu}$ and a line 
element $ds^2=g_{\mu\nu}dx^\mu dx^\nu$. The difference from Einstein's general
relativity is that our coordinates are: $x^0$ is a velocitylike coordinate 
(rather than a timelike coordinate), thus $x^0=\tau v$ where $\tau$ is the
Hubble time in the zero-gravity limit and $v$ the velocity. The coordinate 
$x^0=\tau v$ is the 
comparable to $x^0=ct$ where $c$ is the speed of light and $t$ is the time in
ordinary general relativity. The other three coordinates $x^k$, $k=1,2,3$, are
spacelike, just as in general relativity theory.

An immediate consequence of the above choice of coordinates is that the null
condition $ds=0$ describes the expansion of the Universe in the curved 
spacevelocity (generalized Hubble's law with gravitation) as compared to the 
propagation of light in the curved spacetime in general relativity. This means
one solves the field equations (to be given in the sequel) for the metric 
tensor, then from the null condition $ds=0$ one obtains immedialety the 
dependence of the relative distances between the galaxies on their relative
velocities.

As usual in gravitational theories, one equates geometry to physics. The first 
is expressed by means of a combination of the Ricci tensor and the Ricci
scalar, and follows to be naturally either the Ricci trace-free tensor or the
Einstein tensor. The Ricci trace-free tensor does not fit gravitation in 
general, and
the Einstein tensor is a natural candidate. The physical part is expressed by
the energy-momentum tensor which now has a different physical meaning from 
that in Einstein's theory. More important, the coupling constant that relates 
geometry to physics is now also {\it different}. 

Accordingly the field equations are
$$G_{\mu\nu}=R_{\mu\nu}-g_{\mu\nu}R/2=\kappa T_{\mu\nu},\eqno(3)$$
exactly as in Einstein's theory, with $\kappa$ given by
$\kappa=8\pi k/\tau^4$, (in general relativity it is given by $8\pi G/c^4$),
where $k$ is given by $k=G\tau^2/c^2$, with $G$ being Newton's gravitational
constant, and $\tau$ the Hubble constant time. When the equations of motion 
will be written in terms of velocity instead of time, the constant $k$ will
replace $G$. Using the above equations one then has $\kappa=8\pi G/c^2\tau^2$.

The energy-momentum tensor $T^{\mu\nu}$ is constructed, along the lines of
general relativity  theory, with the speed of light being replaced by the
Hubble constant time. If $\rho$ is the average mass density of the Universe,
then it will be assumed that $T^{\mu\nu}=\rho u^\mu u^\nu,$ where $u^\mu=
dx^\mu/ds$ is the four-velocity.
In general relativity theory one takes $T_0^0=\rho$. In Newtonian gravity one
has the Poisson equation $\nabla^2\phi=4\pi G\rho$. At points where $\rho=0$
one solves the vacuum Einstein field equations in general relativity and the 
Laplace equation 
$\nabla^2\phi=0$ in Newtonian gravity. In both theories a null (zero) solution
is allowed as a trivial case. In cosmology, however, there exists no situation
at which $\rho$ can be zero because the Universe is filled with matter. In
order to be able to have zero on the right-hand side of Eq. (3) one takes 
$T_0^0$ not as equal to $\rho$, but to $\rho_{eff}=\rho-\rho_c$, where 
$\rho_c$ is the critical mass density, 
a {\it constant} in CGR given by $\rho_c=3/8\pi G\tau^2$, whose value is 
$\rho_c\approx 10^{-29}g/cm^3$, a few hydrogen atoms per cubic meter. 
Accordingly one takes
$$T^{\mu\nu}=\rho_{eff}u^\mu u^\nu;\mbox{\hspace{5mm}}\rho_{eff}=\rho-\rho_c
\eqno(4)$$
for the energy-momentum tensor.

In the next sections we apply CGR to obtain the accelerating expanding 
Universe and related subjects.
\section{Gravitational Field Equations}
In the four-dimensional spacevelocity the spherically symmetric metric is 
given by
$$ds^2=\tau^2dv^2-e^\mu dr^2-R^2\left(d\theta^2+\sin^2\theta d\phi^2\right),
\eqno(5)$$
where $\mu$ and $R$ are functions of $v$ and $r$ alone, and comoving 
coordinates $x^\mu=(x^0,x^1,x^2,x^3)=(\tau v,r,\theta,\phi)$ have been used. 
With the above choice of coordinates, the zero-component of the geodesic
equation becomes an identity, and since $r$, $\theta$ and $\phi$ are constants
along the geodesics, one has $dx^0=ds$ and therefore
$$u^\alpha=u_\alpha=\left(1,0,0,0\right).\eqno(6)$$
The metric (5) shows that the area of the sphere $r=constant$ is given by
$4\pi R^2$ and that $R$ should satisfy $R'=\partial R/\partial r>0$. The
possibility that $R'=0$ at a point $r_0$ is excluded since it would
allow the lines $r=constants$ at the neighboring points $r_0$ and $r_0+dr$ to
coincide at $r_0$, thus creating a caustic surface at which the comoving 
coordinates break down.

As has been shown in the previous sections the Universe expands by the null
condition $ds=0$, and if the expansion is spherically symmetric one has
$d\theta=d\phi=0$. The metric (5) then yields
$$\tau^2 dv^2-e^\mu dr^2=0,\eqno(7)$$
thus
$$dr/dv=\tau e^{-\mu/2}.\eqno(8)$$
This is the differential equation that determines the Universe expansion. In
the following we solve the gravitational field equations in order to find out
the function $\mu\left(r.v\right)$.

The gravitational field equations (3), written in the form
$$R_{\mu\nu}=\kappa\left(T_{\mu\nu}-g_{\mu\nu}T/2\right),\eqno(9)$$
where 
$$T_{\mu\nu}=\rho_{eff}u_\mu u_\nu+p\left(u_\mu u_\nu-g_{\mu\nu}\right), 
\eqno(10)$$
with $\rho_{eff}=\rho-\rho_c$ and $T=T_{\mu\nu}g^{\mu\nu}$, are now solved.
Using Eq. (6) one finds that the only nonvanishing components of $T_{\mu\nu}$ 
are $T_{00}=\tau^2\rho_{eff}$, $T_{11}=c^{-1}\tau pe^\mu$, $T_{22}=c^{-1}\tau 
pR^2$ and $T_{33}=c^{-1}\tau pR^2\sin^2\theta$, and that $T=\tau^2\rho_{eff}-
3c^{-1}\tau p$.

The only nonvanishing components of the Ricci tensor yield (dots and primes 
denote differentiation with respect to $v$ and $r$, respectively), using Eq. 
(9), the following field equations:
$$R_{00}=-\ddot{\mu}/2-2\ddot{R}/R-\dot{\mu}^2/4=
\kappa\left(\tau^2\rho_{eff}+3c^{-1}\tau p\right)/2,\eqno(11a)$$
$$R_{01}=R'\dot{\mu}/R-2\dot{R}'/R=0,\eqno(11b)$$
$$R_{11}=e^\mu\left(\frac{1}{2}\ddot{\mu}+\frac{1}{4}\dot{\mu}^2+\frac{1}{R}
\dot{\mu}\dot{R}\right)+\frac{1}{R}\left(\mu'R'-2R''\right)$$
$$=\kappa e^\mu\left(\tau^2\rho_{eff}-c^{-1}\tau p\right)/2,
\eqno(11c)$$
$$R_{22}=R\ddot{R}+\frac{1}{2}R\dot{R}\dot{\mu}+\dot{R}^2+1-e^{-\mu}\left(
RR''-\frac{1}{2}RR'\mu'+R'^2\right)$$
$$=\kappa R^2\left(\tau^2\rho_{eff}-c^{-1}\tau p\right)/2,\eqno(11d)$$
$$R_{33}=\sin^2\theta R_{22}=\kappa R^2\sin^2\theta\left(\tau^2
\rho_{eff}-c^{-1}\tau p\right)/2.\eqno(11e)$$

The field equations obtained for the components 00, 01, 11, and 22 (the 33 
component contributes no new information) are given by
$$-\ddot{\mu}-4\ddot{R}/R-\dot{\mu}^2/2=\kappa\left(
\tau^2\rho_{eff}+3c^{-1}\tau p\right),
\eqno(12)$$
$$2\dot{R}'-R'\dot{\mu}=0, \eqno(13)$$
$$\ddot{\mu}+\frac{1}{2}\dot{\mu}^2+\frac{2}{R}\dot{R}\dot{\mu}+e^{-\mu}\left(
\frac{2}{R}R'\mu'-\frac{4}{R}R''\right)=\kappa\left(\tau^2\rho_{eff}-c^{-1}\tau 
p\right) \eqno(14)$$
$$\frac{2}{R}\ddot{R}+2\left(\frac{\dot{R}}{R}\right)^2+\frac{1}{R}\dot{R}
\dot{\mu}+\frac{2}{R^2}+e^{-\mu}\left[\frac{1}{R}R'\mu'-2\left(\frac{R'}{R}
\right)^2-\frac{2}{R}R''\right]$$
$$=\kappa\left(\tau^2\rho_{eff}-c^{-1}\tau p\right).\eqno(15)$$
It is convenient to eliminate the term with the second velocity-derivative of
$\mu$ from the above equations. This can easily be done, and combinations of 
Eqs. (12)--(15) then give the following set of three independent field 
equations:
$$e^\mu\left(2R\ddot{R}+\dot{R}^2+1\right)-R'^2=-\kappa\tau c^{-1} e^\mu R^2p,
\eqno(16)$$
$$2\dot{R}'-R'\dot{\mu}=0, \eqno(17)$$
$$e^{-\mu}\left[\frac{1}{R}R'\mu'-\left(\frac{R'}{R}\right)^2-\frac{2}{R}R''
\right]+\frac{1}{R}\dot{R}\dot{\mu}+\left(\frac{\dot{R}}{R}\right)^2+
\frac{1}{R^2}$$
$$=\kappa\tau^2\rho_{eff}, \eqno(18)$$
other equations being trivial combinations of (16)--(18).
\section{Solution of the Field Equations}
The solution of Eq. (17) satisfying the condition $R'>0$ is given by
$$e^\mu=R'^2/\left[1+f\left(r\right)\right],\eqno(19)$$
where $f\left(r\right)$ is an arbitrary function of the coordinate $r$ and 
satisfies the
condition $f\left(r\right)+1>0$. Substituting (19) in the other two field
equations (16) and (18) then gives
$$2R\ddot{R}+\dot{R}^2-f=-\kappa c^{-1}\tau R^2p, \eqno(20)$$
$$\left(2\dot{R}\dot{R'}-f'\right)/\left(RR'\right)+\left(\dot{R}^2-f
\right)/R^2=\kappa\tau^2\rho_{eff},\eqno(21)$$
respectively.

The simplest solution of the above two equations, which satisfies the 
condition $R'=1>0$, is given by
$$R=r.\eqno(22)$$
Using Eq. (22) in Eqs. (20) and (21) gives
$$f\left(r\right)=\kappa c^{-1}\tau pr^2,\eqno(23)$$
and
$$f'+f/r=-\kappa\tau^2\rho_{eff}r,\eqno(24)$$
respectively. The solution of Eq. (24) is the sum of the solutions of the
homogeneous equation
$$f'+f/r=0,\eqno(25)$$
and a particular solution of Eq. (24). These are given by
$$f_1=-2Gm/(c^2r),\eqno(26)$$
and 
$$f_2=-\kappa\tau^2\rho_{eff}r^2/3.\eqno(27)$$

The solution $f_1$ represents a particle at the origin of coordinates and as
such is not relevant to our problem. We take, accordingly, $f_2$ as the 
general solution,
$$f\left(r\right)=-\frac{\kappa}{3}\tau^2\rho_{eff}r^2=-\frac{\kappa}{3}\tau^2
\left(\rho-\rho_c\right)r^2$$
$$=-\frac{\kappa}{3}\tau^2\rho_c\left(
\rho/\rho_c-1\right)r^2.\eqno(28)$$
Using the values of $\kappa=8\pi G/c^2\tau^2$ and $\rho_c=3/8\pi G\tau^2$, we
obtain
$$f\left(r\right)=\frac{1-\Omega_m}{c^2\tau^2}r^2,\eqno(29)$$
where $\Omega_m=\rho/\rho_c$.

The two solutions given by Eqs. (23) and (29) for $f(r)$ can now be 
equated, giving
$$p=\frac{1-\Omega_m}{\kappa c\tau^3}=\frac{c}{\tau}\frac{1-\Omega_m}{8\pi G}
=4.544\left(1-\Omega_m\right)\times 10^{-2} g/cm^2.\eqno(30)$$
Furthermore, from Eqs. (19) and (22) we find that 
$$e^{-\mu}=1+f\left(r\right)=1+\tau c^{-1}\kappa pr^2=1+
\frac{1-\Omega_m}{c^2\tau^2}r^2.\eqno(31)$$

It  will be recalled that the Universe expansion is determined by Eq. (8),
$dr/dv=\tau e^{-\mu/2}$. The only thing that is left to be determined is the
signs of $(1-\Omega_m)$ or the pressure $p$.

Thus we have
$$\frac{dr}{dv}=\tau\sqrt{1+\kappa\tau c^{-1}pr^2}=\tau\sqrt{1+
\frac{1-\Omega_m}{c^2\tau^2}r^2}.\eqno(32)$$
For simplicity we confine ourselves to the linear approximation, thus Eq. 
(32) yields
$$\frac{dr}{dv}=\tau\left(1+\frac{\kappa}{2}\tau c^{-1}pr^2\right)=
\tau\left[1+\frac{1-\Omega_m}{2c^2\tau^2}r^2\right].\eqno(33)$$

The second term in the square bracket in the above equation represents the
deviation due to gravity from the standard Hubble law. For without that term,
Eq. (33) reduces to $dr/dv=\tau$, thus $r=\tau v+const$. The constant can be 
taken zero if one assumes, as usual, that at $r=0$ the velocity should also
vanish. Thus $r=\tau v$, or $v=H_0r$ (since $H_0\approx 1/\tau$). Accordingly, 
the equation of motion (33) describes the expansion of the Universe when 
$\Omega_m=1$, namely when $\rho=\rho_c$. The equation then coincides with the 
standard Hubble law. We will not go through that and the reader is referred to
Appendix A of Ref. 1.
\section{Physical Meaning}
To see the physical meaning of these solutions, however, one does not need the
exact solutions. Rather, it is enough to write down the solutions in the
lowest approximation in $\tau^{-1}$. One obtains, by differentiating Eq. 
(33) with respect to $v$, for $\Omega_m>1$,
$$d^2r/dv^2=-kr;\mbox{\hspace{10mm}}k=\left(\Omega_m-1\right)/\left(2c^2\right),\eqno(34)$$
the solution of which is 
$$r\left(v\right)=A\sin\alpha\frac{v}{c}+B\cos\alpha\frac{v}{c},\eqno(35)$$
where $\alpha^2=(\Omega_m-1)/2$ and $A$ and $B$ are constants. The latter can be
determined by the initial condition $r\left(0\right)=0=B$ and $dr\left(0
\right)/dv=\tau=A\alpha/c$, thus
$$r\left(v\right)=\frac{c\tau}{\alpha}\sin\alpha\frac{v}{c}.\eqno(36)$$
This is obviously a closed Universe, and presents a decelerating expansion.

For $\Omega_m<1$ we have
$$d^2r/dv^2=\frac{\left(1-\Omega_m\right)r}{2c^2},\eqno(37)$$
whose solution, using the same initial conditions, is
$$r\left(v\right)=\frac{c\tau}{\beta}\sinh\beta\frac{v}{c},\eqno(38)$$
where $\beta^2=(1-\Omega_m)/2$. This is now an open accelerating Universe.

For $\Omega_m=1$ we have, of course, $r=\tau v$.
\section{The Accelerating Universe}
We finally determine which of the three cases of expansion is the one at 
present epoch of time. To this end we have to write the solutions (36) and 
(38) in ordinary Hubble's law form $v=H_0r$. Expanding Eqs. (36) and 
(38) in power series in $v/c$ and keeping terms up to the second order, we
obtain
$$r=\tau v\left(1-\alpha^2v^2/6c^2\right), \hspace{5mm} 
r=\tau v\left(1+\beta^2v^2/6c^2\right), \eqno(39)$$
for $\Omega_m>1$ and $\Omega_m<1$, respectively. Using now the expressions for 
$\alpha$ and $\beta$, Eqs. (39) then reduce into the single equation
$$r=\tau v\left[1+\left(1-\Omega_m\right)v^2/6c^2\right].\eqno(40)$$
Inverting now this equation by writing it as $v=H_0r$, we obtain in the lowest
approximation
$$H_0=h\left[1-\left(1-\Omega_m\right)v^2/6c^2\right],\eqno(41)$$
where $h=\tau^{-1}$. To the same approximation one also obtains
$$H_0=h\left[1-\left(1-\Omega_m\right)z^2/6\right]=h\left[1-\left(1-\Omega_m
\right)r^2/6c^2\tau^2\right],\eqno(42)$$
where $z$ is the redshift parameter.
As is seen, and it is confirmed by experiments, $H_0$ depends on the distance 
it is being measured; it has physical meaning only at the zero-distance limit,
namely when measured {\it locally}, in which case it becomes $h=1/\tau$.

It follows that the measured value of $H_0$ depends on the ``short"
and ``long" distance scales \cite{8}. The farther the distance $H_0$ is being 
measured, the lower the value for $H_0$ is obtained. By Eq. (42) this is
possible only when $\Omega_m<1$, namely when the Universe is accelerating. By
Eq. (30) we also find that the pressure is positive. 

The possibility that the Universe expansion is accelerating was first 
predicted using CGR by the author in 1996 [2] before the supernovae 
experiments results became known.

It  will be noted that the constant expansion is just a transition stage 
between the decelerating and the accelerating expansions as the Universe
evolves toward its present situation.
\begin{figure}
\centerline{\psfig{file=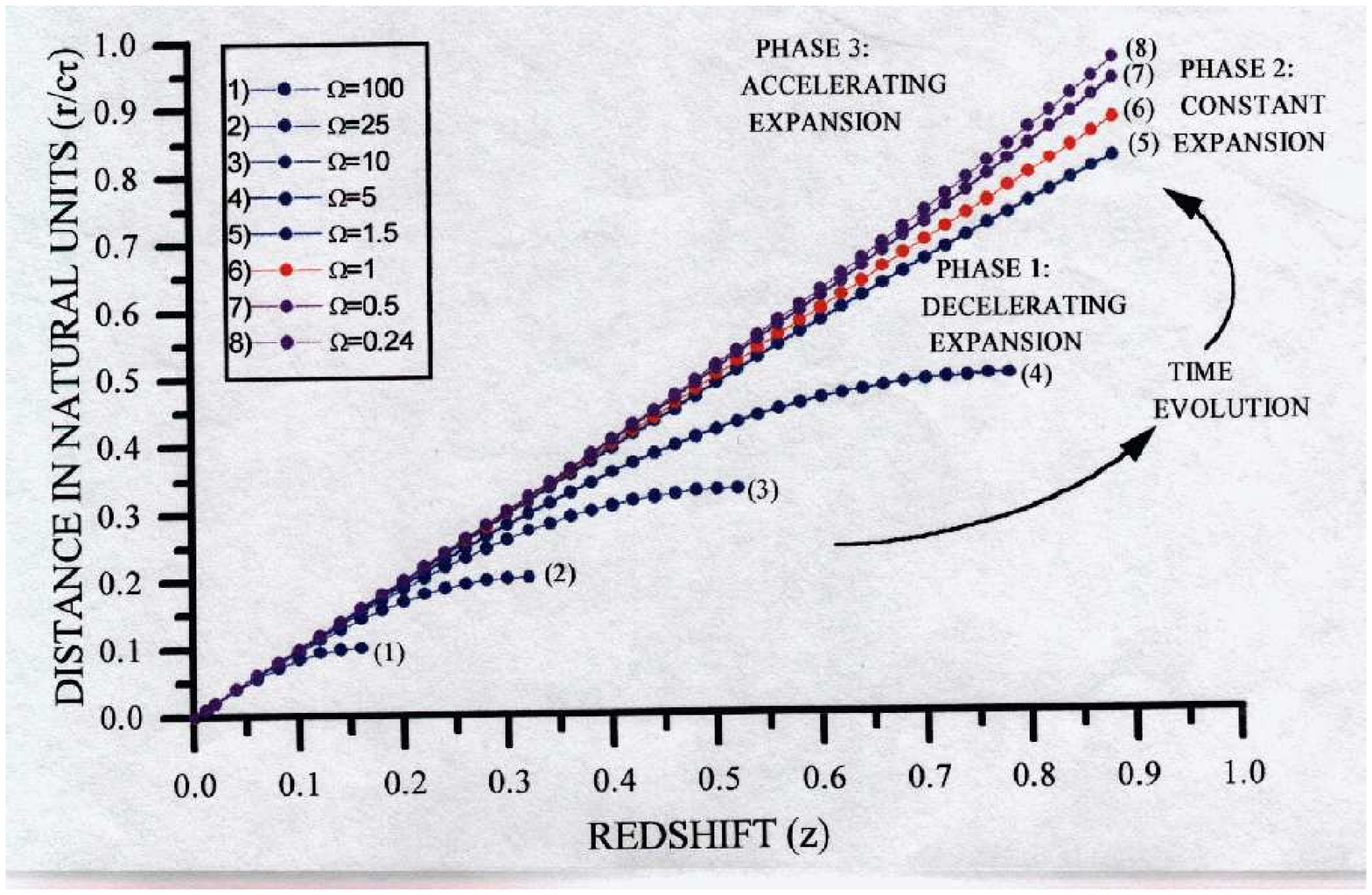,angle=180,width=5cm}}
\vspace*{8pt}
\caption{}
\end{figure}

Figure 1 (Behar and Carmeli) describes the Hubble diagram of the above solutions for the three 
types of expansion for values of $\Omega_m$ from 100 to 0.245. The figure
describes the three-phase evolution of the Universe. Curves (1)-(5) represent
the stages of {\it decelerating expansion} according to Eq. (36). As the density 
of matter $\rho$ decreases, the Universe goes over from the lower curves to 
the upper ones, but it does not have enough time to close up to a big crunch.
The Universe subsequently goes over to curve (6) with $\Omega_m=1$, at which time
it has a constant expansion for a fraction of a second. This then followed
by going to the upper curves (7) and (8) with $\Omega_m<1$, where the Universe
expands with {\it acceleration} according to Eq. (38). Curve no. 8 fits
the present situation of the Universe. For curves (1)-(4) in the diagram we
use the cutoff when the curves were at their maximum. In Table 1 we present 
the cosmic times with respect to the big bang, the cosmic radiation
temperature and the pressure for each of the curves in Fig. 1.
Figure 2 (Riess {\it et al.}) shows the experimental results.
\begin{figure}
\centerline{\psfig{file=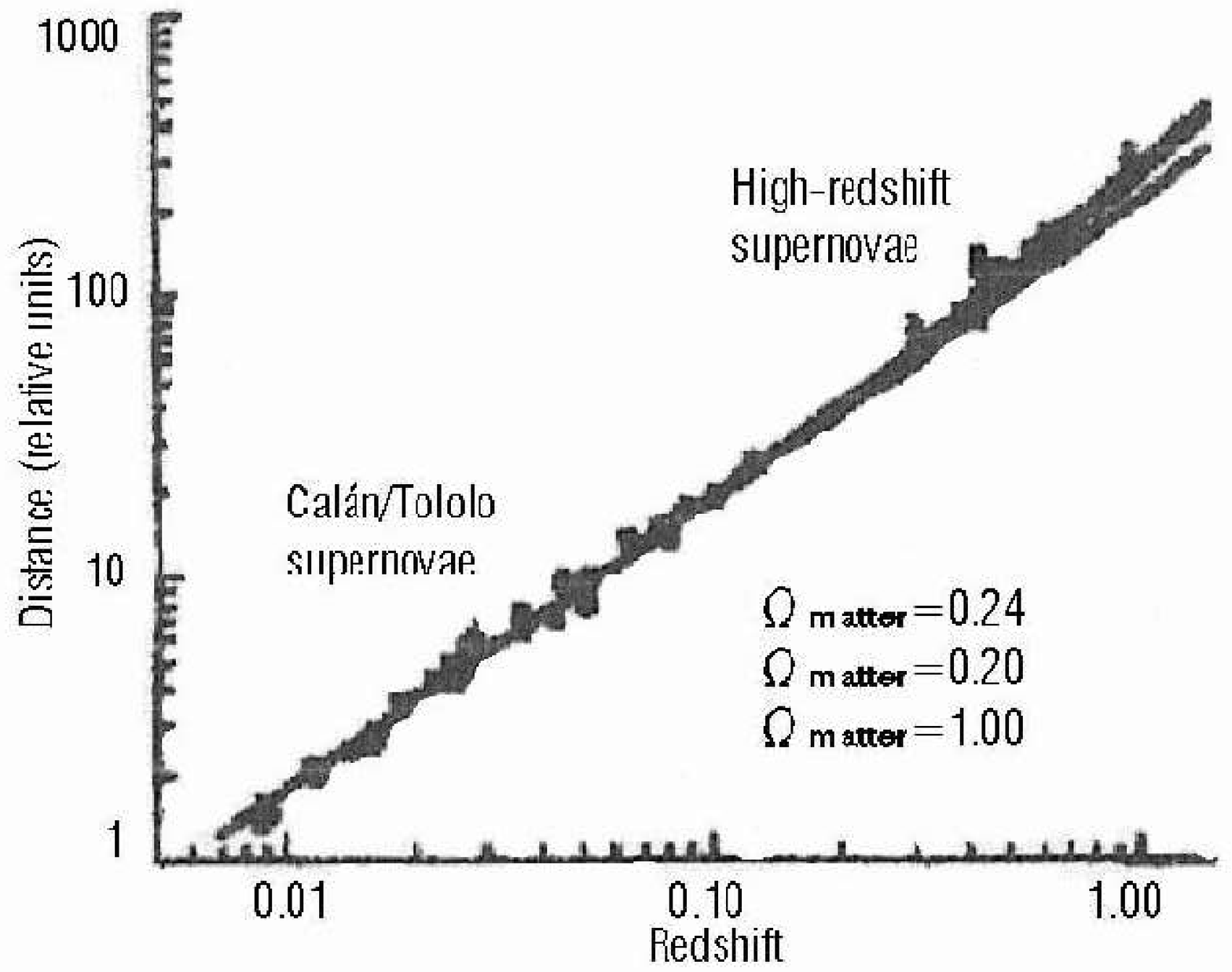,angle=-90,width=5cm}}
\vspace*{8pt}
\caption{}
\end{figure}
\begin{table}[h]
\tbl{The Cosmic Times with respect to the Big Bang, the Cosmic
Temperature and the Cosmic Pressure for each of the Curves in Fig. 1.}
{\begin{tabular}{c r@{}l c r@{.}l r@{}l r@{}l}
\hline\\ 
Curve&\multicolumn{2}{c}{$\Omega_m$}&Time in Units &\multicolumn{2}{c}{Time}&
\multicolumn{2}{c}{Temperature}&\multicolumn{2}{c}{Pressure}\\
No$^\star$.& & & of $\tau$ &\multicolumn{2}{c}{(Gyr)}& &(K)&
\multicolumn{2}{c}{(g/cm$^2$)}\\
\hline\\
\multicolumn{10}{c}{DECELERATING EXPANSION}\\
1&100& &$3.1\times 10^{-6}$&$3$&$87\times 10^{-5}$&1096&&-&4.499\\
2&25& &$9.8\times 10^{-5}$&$1$&$22\times 10^{-3}$&195&.0&-&1.091\\
3&10& &$3.0\times 10^{-4}$&$3$&$75\times 10^{-3}$&111&.5&-&0.409\\
4&5& &$1.2\times 10^{-3}$&$1$&$50\times 10^{-2}$&58&.20&-&0.182\\
5&1&.5&$1.3\times 10^{-2}$&$1$&$62\times 10^{-1}$&16&.43&-&0.023\\
\multicolumn{10}{c}{CONSTANT EXPANSION}\\ 
6&1& &$3.0\times 10^{-2}$&$3$&$75\times 10^{-1}$&11&.15&&0\\
\multicolumn{10}{c}{ACCELERATING EXPANSION}\\ 
7&0&.5&$1.3\times 10^{-1}$&$1$&$62$&5&.538&+&0.023\\
8&0&.245&$1.0$&$12$&$50$&2&.730&+&0.034\\
\hline\\ 
\end{tabular}}
\end{table}

To summarize, a theory of cosmology has been presented in which the dynamical
variables are those of Hubble, i.e. distances and velocities. The theory
describes the Universe as having a three-phase evolution with a decelerating 
expansion, followed by a constant and an accelerating expansion, and it
predicts that the Universe is now in the latter phase. As the density of 
matter decreases, while the Universe is at the decelerating phase, it does not
have enough time to close up to a big crunch. Rather, it goes to the 
constant-expansion phase, and then to the accelerating stage. As we have seen,
the equation obtained for the Universe expansion, Eq. (38), is very simple.

\section{Theory versus Experiment}
The Einstein gravitational 
field equations  
with the added cosmological term are \cite{9}:
$$R_{\mu\nu}-g_{\mu\nu}R/2+\Lambda g_{\mu\nu}=\kappa T_{\mu\nu},
\eqno(43)$$
where $\Lambda$ is the cosmological constant, the value of which is supposed to
be determined by experiment. In Eq. (43) $R_{\mu\nu}$ and $R$ are the Ricci 
tensor and scalar, respectively, $\kappa=8\pi G$, where $G$ is Newton's constant
and the speed of light is taken as unity.

Recently the two groups (the {\it Supernovae Cosmology Project} and the {\it 
High-Z Supernova Team}) concluded that the expansion of the Universe is 
accelerating \cite{10,11,12,13,14,15,16}. The two groups had discovered and measured moderately 
high redshift ($0.3<z<0.9$) supernovae, and found that they were fainter than
what one would expect them to be if the cosmos expansion were slowing down or
constant. Both teams obtained
$\Omega_m\approx 0.3$, $\Omega_\Lambda\approx 0.7$,
and ruled out the traditional ($\Omega_m$, $\Omega_\Lambda$)=(1, 0)
Universe. Their value for the density parameter $\Omega_\Lambda$ corresponds to
a cosmological constant that is small but, nevertheless, nonzero and positive,
$\Lambda\approx 10^{-52}\mbox{\rm m}^{-2}\approx 10^{-35}\mbox{\rm s}^{-2}.$

In previous sections a four-dimensional cosmological theory (CGR) was 
presented. Although the theory has no cosmological constant, it predicts that 
the Universe accelerates and hence it has the equivalence of a positive 
cosmological constant in 
Einstein's general relativity. In the framework of this theory (see Section 
2) the 
zero-zero component of the field equations (3) is written as
$$R_0^0-\frac{1}{2}\delta_0^0R=\kappa\rho_{eff}=\kappa\left(\rho-\rho_c
\right),\eqno(44)$$
where $\rho_c=3/\kappa\tau^2$ is the critical mass density
and $\tau$ is Hubble's time in the zero-gravity limit.

Comparing Eq. (44) with the zero-zero component of Eq. (43), one obtains 
the expression for the cosmological constant of general relativity,
$\Lambda=\kappa\rho_c=3/\tau^2.$

To find out the numerical value of $\tau$ we use the relationship between
$h=\tau^{-1}$ and $H_0$ given by Eq. (42) (CR denote values according to
Cosmological Relativity): 
$$H_0=h\left[1-\left(1-\Omega_m^{CR}\right)z^2/6\right],\eqno(45)$$
where $z=v/c$ is the redshift and $\Omega_m^{CR}=\rho_m/\rho_c$ with 
$\rho_c=3h^2/8\pi G$. (Notice that our $\rho_c=1.194\times 10^{-29}g/cm^3$ is different from 
the standard $\rho_c$ defined with $H_0$.) The redshift parameter $z$ 
determines the distance at which $H_0$ is measured. We choose $z=1$ and take 
for 
$\Omega_m^{CR}=0.245,$
its value at the present time (see Table 1) (corresponds to 0.32 in the 
standard theory), Eq. (45) then gives
$H_0=0.874h.$
At the value $z=1$ the corresponding Hubble parameter $H_0$ according to the 
latest results from HST can be taken \cite{17} as $H_0=70$km/s-Mpc, thus 
$h=(70/0.874)$km/s-Mpc, or
$h=80.092\mbox{\rm km/s-Mpc},$
and 
$\tau=12.486 Gyr=3.938\times 10^{17}s.$
\begin{table}[h]
\tbl{Cosmological parameters in cosmological general  relativity and in
standard theory}
{\begin{tabular}{p{35mm}p{35mm}p{35mm}}
\hline\\
&COSMOLOGICAL&STANDARD\\
&RELATIVITY&THEORY\\
\hline\\
Theory type&Spacevelocity&Spacetime\\
Expansion&Tri-phase:&One phase\\
type&decelerating, constant,&\\
&accelerating&\\
Present expansion&Accelerating&One of three\\
&(predicted)&possibilities\\
Pressure&$0.034g/cm^2$&Negative\\
Cosmological constant&$1.934\times 10^{-35}s^{-2}$&Depends\\
&(predicted)&\\
$\Omega_T=\Omega_m+\Omega_\Lambda$&1.009&Depends\\
Constant-expansion&8.5Gyr ago&No prediction\\
occurs at&(adjusted to gravity)&\\
Constant-expansion&Fraction of&Not known\\
duration&second&\\
Temperature at&146K&No prediction\\
constant expansion&(adjusted to gravity)&\\
\hline
\end{tabular}}
\end{table}

What is left is to find the value of $\Omega_\Lambda^{CR}$. We have 
$\Omega_\Lambda^{CR}=\rho_c^{ST}/\rho_c$, where $\rho_c^{ST}=3H_0^2/8\pi 
G$ and $\rho_c=3h^2/8\pi G$. Thus $\Omega_\Lambda^{CR}=(H_0/h)^2=0.874^2$,
or
$\Omega_\Lambda^{CR}=0.764.$
As is seen from the above one has 
$$\Omega_T=\Omega_m^{CR}+\Omega_\Lambda^{CR}=0.245+0.764=1.009\approx 1,
\eqno(46)$$
which means the Universe is Euclidean.

As a final result we calculate the cosmological constant in the form
$$\Lambda=3/\tau^2=1.934\times 10^{-35}s^{-2}.\eqno(47)$$

Our results confirm those of the supernovae experiments and indicate on the
existance of the dark energy as has recently received confirmation from the
Boomerang cosmic microwave background experiment \cite{18,19}, which showed that 
the Universe is Euclidean.
\section{Concluding Remarks}
In this paper the cosmological general relativity, a relativistic theory in
spacevelocity, has been presented and applied to the problem of the expansion 
of the Universe. The theory, which predicts a positive pressure for the 
Universe now, describes the Universe as having a three-phase
evolution: decelerating, constant and accelerating expansion, but it is now in
the latter stage. Furthermore, the cosmological constant that was
extracted from the theory agrees with the experimental result. Finally, it has
also been shown that the three-dimensional spatial space of the Universe is
Euclidean, again in agreement with observations. 
\begin{figure}
\centerline{\psfig{file=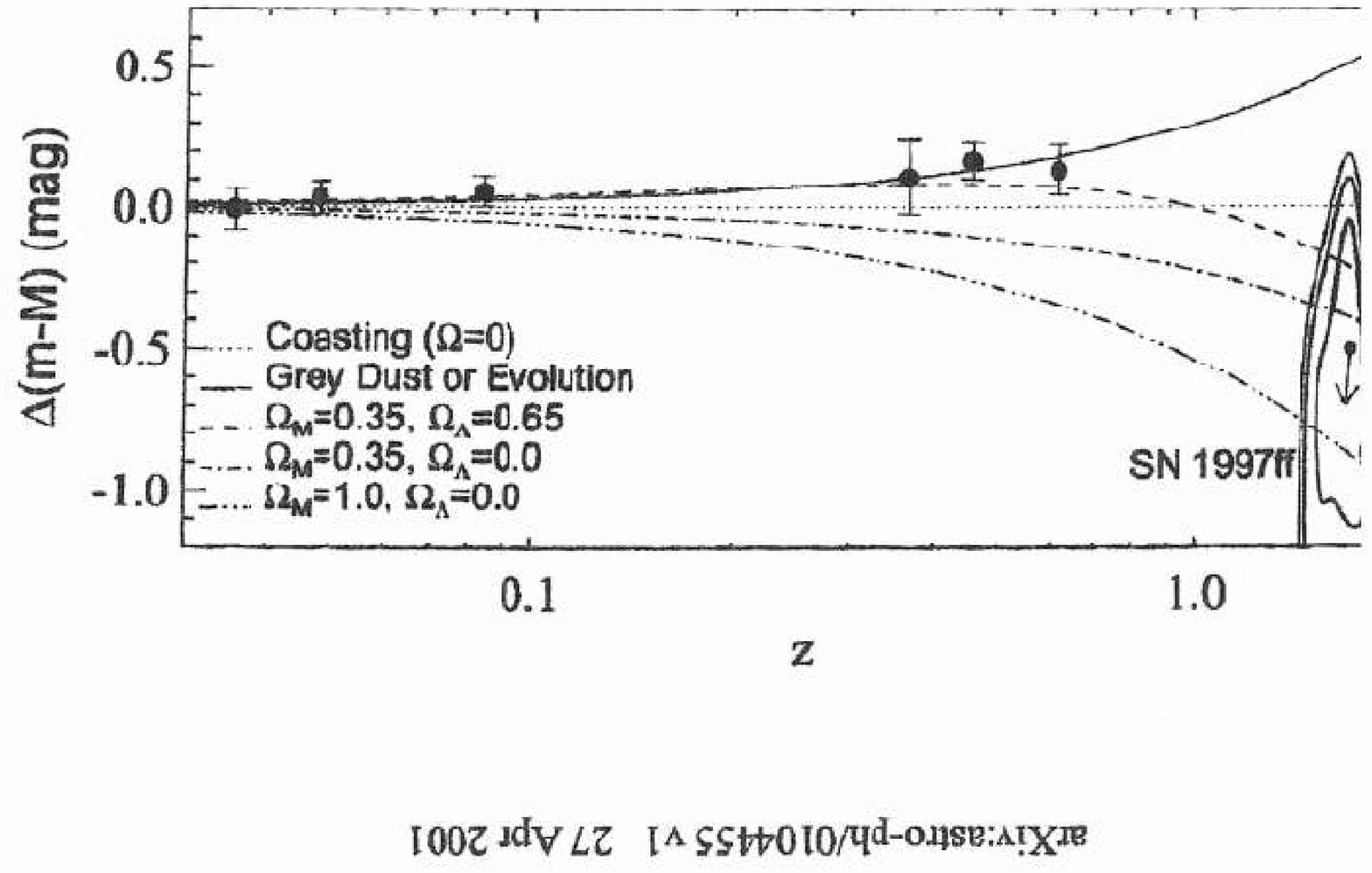,angle=90,width=8cm}}
\vspace*{8pt}
\caption{}
\end{figure}

Recently \cite{20,21}, more confirmation to the Universe accelerating expansion 
came from 
the most distant supernova, SN 1997ff, that was recorded by the Hubble Space
Telescope. As has been pointed out before, if we look back far enough, we 
should find a decelerating expansion (curves 1-5 in Figure 1). Beyond $z=1$
one should see an earlier time when the mass density was dominant. The 
measurements obtained from SN 1997ff's redshift and brightness provide a 
direct proof for the transition from past decelerating to present 
accelerating expansion (see Figure 3, Riess {\it et  al.}). The measurements 
also exclude the possibility that
the acceleration of the Universe is not real but is due to other astrophysical 
effects such as dust.

Table 2 gives some of the cosmological parameters obtained here and in the
standard theory.

\end{document}